\documentstyle[amsbsy]{elsart}
\begin{document}
\newcommand{\re}{\mathop{\mathrm{Re}}}
\newcommand{\im}{\mathop{\mathrm{Im}}}
\begin{frontmatter}
\title{
T-odd particle spin rotation in
nuclear target}
\author{S.L. Cherkas }
\address{
Institute of Nuclear Problems, Bobruiskaya 11, Minsk 220050, Belarus}
\begin{abstract}
It is shown that
particle spin rotation in nuclear target, caused by
the imaginary part of the spin-dependent forward scattering amplitude,
is T-odd rotation.
\end{abstract}
\end{frontmatter}

Prediction of the effect of neutron spin rotation \cite{bar}
in matter with polarized
nuclei target has
initiated
a  research into polarization
effects arising at passing of a particle with spin through
nuclear target \cite {bar1}. Phenomenon of
nuclear spin precision  was experimentally confirmed \cite {abr0,fort}.
As it is known \cite {bar,abr,fort} a frequency of spin rotation is
proportional to the real part of the forward spin-dependent
scattering amplitude,
that allows to consider spin rotation, as a
method of measurement of this quantity.

The influence of the imaginary part of the amplitude on spin rotation
was also considered \cite {bar,abr,fort}, however, it was not
clearly shown anywhere, that for the completely polarized beam
the imaginary part of the spin-dependent forward
scattering amplitude
results in T-odd spin rotation.

Consider motion of
$\frac{1}{2}$
spin  particle in the polarized
substance.
The dependence of the spin wavefunction of the particle on time is
written down as \cite {bar1}
\begin {equation}
\psi (t) = \exp\left (i\hat n k v t\right) \psi_0 ~~,
\end {equation}
where $ \psi_0 $
is the initial spin wavefunction of the particle
at a medium entry,
$k,v $ is particle
wave vector and
speed respectively,
$ \hat n $ is effective refraction index of a particle inside
substance depending on spin:
\begin {equation}
\hat n=1 + \frac {2\pi\rho} {k^2} \hat f (0).
\end {equation}
Here $ \rho $ is the density of scatterers in the matter,
$ \hat f (0) $ is
the amplitude of particle zero-angle elastic scattering by
a scattering centre of the medium.
Forward scattering
amplitude can be represented as
$ \hat f (0) = A_0+A_s (\boldsymbol \sigma \cdot \boldsymbol J) ~ $, where
$ \boldsymbol \sigma $ is Pauli matrix,
$ \boldsymbol J $ is direction of  the
polarization of the scatterers in the medium.
In general case
$ \boldsymbol J $ represents a vector made of every possible combinations
of polarization,
qudroupole polarization
of the scatterers in medium, wave vector
of the incident particle \cite {bar1}.
$\boldsymbol J$ can contain P-odd, T-odd and
P,T- odd combinations.
Knowing the $\psi(t)$ polarization evolution can be
obtained:
\begin{equation}
O(t)=<\psi^+(t)\mid \boldsymbol \sigma \mid \psi(t)>
<\psi^+(t)\mid\psi(t)>^{-1} ~.
\end{equation}
In works \cite{bar,abr} $O(t)$ has founded for the particular case
when initial polarization is directed perpendicular to the
$\boldsymbol J$.
\begin{eqnarray}
O_x(t)=\cos\left(\frac{4\pi\rho v }{k} \re A_s t\right)
/\cosh\left( \frac{4\pi\rho v}{k} \im A_s t  \right)
\nonumber \\
O_y(t)=-\sin\left(\frac{4\pi\rho v }{k} \re A_s t\right)
/\cosh\left( \frac{4\pi\rho v}{k} \im A_s  t \right)
\nonumber \\
O_z(t)=-\tanh\left(\frac{4\pi\rho v}{k} \im A_s t \right)
\end{eqnarray}
When $\im A_s=0$ polarization motion $\boldsymbol O(t)$
is pure rotation around
$\boldsymbol J$. In the case of absorbtion the vector
$\boldsymbol O(t)$ moves on sphere
suffice because
$\boldsymbol O\cdot \boldsymbol O=1$.

Let's
consider this motion in details
on a bases of spin
density matrix of a
particle $ \wp (t) = \psi (t) \psi ^ + (t) $.
Under motion of a
particle in matter spin matrix evolution is
described by the equation
\begin {equation}
\frac {\d\wp
(t)}{\d t} = \frac {\d \psi (t)} {\d t} \psi ^ + (t) + \psi (t)
\frac {\d
\psi ^ + (t)} {\d t} =ikv [\hat n, \wp] ~~, \end {equation}
where the
square brackets denote operator commutator.
Representing
the density spin matrix as
$\wp(t)=\frac{\zeta_0(t)+\boldsymbol\sigma\cdot\boldsymbol\zeta(t)}{2}~$,
we find
\begin {eqnarray}
\frac {\d \boldsymbol\zeta} {\d t} = \frac {4\pi\rho v}{k}
\left\{-\im  A_0 \boldsymbol\zeta +\re A_s (\boldsymbol\zeta\times
\boldsymbol J) -\im A_s \zeta_0 \boldsymbol J\right\} \\ \nonumber
\frac {\d \zeta_0} {\d t} = \frac {4\pi\rho v}{k}
\left\{-\im  A_0\zeta_0-\im A_s (\boldsymbol \zeta\cdot \boldsymbol J)
\right\} ~.
\label{33}
\end {eqnarray}
The density matrix is not normalized and vector $ \boldsymbol
\zeta $ is not  a true polarization. True polarization
is
$ \boldsymbol O =\boldsymbol \zeta/\zeta_0 $.
The equation
for $\frac{\d \boldsymbol O}{\d t} $ is
\begin {eqnarray} \frac {\d \boldsymbol O}{\d t}=
\frac{1}{\zeta_0}\frac{\d\boldsymbol
\zeta} {\d t} - \frac {\boldsymbol
\zeta} {\zeta_0^2} \frac {\d\zeta_0} {\d t} ~, ~~~~~~~~~~~~~~~~~~~~
\nonumber \\ \frac {\d \boldsymbol O} {\d t} = \frac {4\pi\rho
v} {k} \left \{\re A_s ~ \boldsymbol O\times \boldsymbol J -\im  A_s
(\boldsymbol J-\boldsymbol O (\boldsymbol J\cdot \boldsymbol O)) \right \}~.
\label {t}
\end {eqnarray}
For completely polarized beam ($ \boldsymbol O\cdot
\boldsymbol O=1 )$
rewrite (\ref {t}) as:
\begin {equation}
\frac {\d \boldsymbol O} {\d t} = \frac {4\pi\rho v} {k} \left
\{\re A_s ~
\boldsymbol O\times \boldsymbol J
+\im  A_s ~\boldsymbol O\times (\boldsymbol O\times \boldsymbol
J) \right \} ~.
\end {equation}
Thus we see, that the imaginary part
of the spin dependent forward scattering
amplitude results in
the T-odd spin rotation
($ \boldsymbol J, \boldsymbol O $ are P-even,
T - odd vectors) around the driven axe.

In this case it is easy to distinguish this "dissipative"
terms
from the "true" T-odd terms caused
by breaking of the time-reversal invariance on
fundamental level,
as the first are terms of the
second order on polarization.
Note, that the similar "dissipative" T-odd terms
of the second order on polarization
appear in the
Bargmann-Michel-Telegdi
equation, describing spin motion
in an external electromagnetic field if to take into account
influence of radiation reaction
on spin motion \cite {rowe}.

We have shown, that
spin rotation in a nuclear target,
proportional to the imaginary part
of the spin-dependent forward scattering amplitude
is a T-odd one.
This rotation can be used for absolute
calibration of the equipment on
spin rotation angle
measurements,
as the imaginary part of the forward scattering amplitude
can be measured through total section, that allows to
calculate
expected rotational angle precisely.

The author wishes to thank Prof. V. Baryshevsky and Dr. K. Batrakov 
for the discassion.  

\end {document}